\documentclass[showpacs,aps,prd,nofootinbib,floatfix,amsmath,amssymb]{revtex4}
\usepackage{graphicx}
\newcommand{\newc}{\newcommand}
\newc{\gsim}{\lower.7ex\hbox{$\;\stackrel{\textstyle>}{\sim}\;$}}
\newc{\lsim}{\lower.7ex\hbox{$\;\stackrel{\textstyle<}{\sim}\;$}}
\begin{document}
\makeatletter
%Feynman slash
\newbox\slashbox \setbox\slashbox=\hbox{$/$}
\newbox\Slashbox \setbox\Slashbox=\hbox{\large$/$}
\def\pFMslash#1{\setbox\@tempboxa=\hbox{$#1$}
  \@tempdima=0.5\wd\slashbox \advance\@tempdima 0.5\wd\@tempboxa
  \copy\slashbox \kern-\@tempdima \box\@tempboxa}
\def\pFMSlash#1{\setbox\@tempboxa=\hbox{$#1$}
  \@tempdima=0.5\wd\Slashbox \advance\@tempdima 0.5\wd\@tempboxa
  \copy\Slashbox \kern-\@tempdima \box\@tempboxa}
\def\FMslash{\protect\pFMslash}
\def\FMSlash{\protect\pFMSlash}
\def\miss#1{\ifmmode{/\mkern-11mu #1}\else{${/\mkern-11mu #1}$}\fi}
%%%% Uso:  \pFMSlash{p}
\makeatother
%\tightenline

\title{Effects of physics beyond the standard model on the neutrino charge radius: an effective Lagrangian approach}
\author{H. Novales--S\' anchez$^{(a)}$, A. Rosado$^{(b)}$, V. Santiago--Ol\' an$^{(b)}$, J. J. Toscano$^{(a,c)}$}
\address{$^{(a)}$Facultad de Ciencias F\'{\i}sico Matem\'aticas,
Benem\'erita Universidad Aut\'onoma de Puebla, Apartado Postal
1152, Puebla, Pue., M\'exico.\\
$^{(b)}$Instituto de F\'{\i}sica, Benem\'erita Universidad
Aut\'onoma de Puebla. Apdo. Postal J-48, C.P. 72570 Puebla, Pue.,
M\'exico.\\
$^{(c)}$Instituto de F\'{\i}sica y Matem\' aticas, Universidad
Michoacana de San Nicol\' as de Hidalgo, Edificio C-3, Ciudad
Universitaria, C.P. 58040, Morelia, Michoac\' an, M\' exico.}
%\author{A. Rosado}
%\address{Instituto de F\'{\i}sica, BUAP. Apdo. Postal J-48, C.P.
%12570 Puebla, Pue., M\'exico}
\date{\today}
%\maketitle
\begin{abstract}
In this work, we look for possible new physics effects on the
electromagnetic charge and anapole form factors, $f_Q(q^2)$ and
$f_A(q^2)$, for a massless Dirac neutrino, when these quantities
are calculated in the context of an effective electroweak
Yang-Mills theory, which induces the most general
$SU_L(2)$--invariant Lorentz tensor structure of nonrenormalizable
type for the $WW\gamma$ vertex. It is found that in this context,
besides the standard model contribution, the additional
contribution to $f_{Q}(q^2)$ and $f_{A}(q^2)$ ($f_{Q}^{O_W}(q^2)$
and $f_{A}^{O_W}(q^2)$, respectively) are gauge independent and
finite functions of $q^2$ after adopting a renormalization scheme.
These form factors, $f_{Q}^{O_W}(q^2)$ and $f_{A}^{O_W}(q^2)$, get
contribution at the one loop level only from the proper neutrino
electromagnetic vertex. Besides, the relation
$f_{Q}^{eff}(q^2)=q^2f_{A}^{eff}(q^2)$
($f_{Q}^{eff}(q^2)=f_{Q}^{SM}(q^2)+f_{Q}^{O_W}(q^2)$,
$f_{A}^{eff}(q^2)=f_{A}^{SM}(q^2)+f_{A}^{O_W}(q^2)$) is still
fulfilled and hence the relation $a_{\nu}^{eff} = \langle
r^2_{\nu} \rangle ^{eff} /6$ ($a_{\nu}^{eff} = a_{\nu}^{SM}+
a_{\nu}^{O_W}$, $\langle r^2_{\nu} \rangle ^{eff} = \langle
r^2_{\nu} \rangle ^{SM}+\langle r^2_{\nu} \rangle ^{O_W}$)is
gotten, just as in the SM. Using the experimental constraint on
the anomalous $WW\gamma$ vertex, a value for the additional
contribution to the charge radius of $|\langle r^2_{\nu}
\rangle^{O_W}| \lsim 10^{-34}\ cm^2$ is obtained, which is one
order of magnitude lower than the SM value.
\end{abstract}
\pacs{13.15.+g, 13.40.Gp, 23.40.Bw, 25.30.Pt}

\maketitle

\setcounter{footnote}{0} \setcounter{page}{1}
\setcounter{section}{0} \setcounter{subsection}{0}
\setcounter{subsubsection}{0}

%%%%%%%%%%%%%%%%%%%%%%%%%%%%%%%%%%%%%%%%%%%%%%%%%%%%%%%%%%%%%%%%%%%%%%%
\section{Introduction.}

Many authors have studied the neutrino charge radius (NCR)
\cite{Bernstein,bardeen,Lee:1977tib,Kim,Kim2,Bardin:1973mg,Aubrecht:1974cg,Monyonko:1984gb,Vogel:1989iv,Ng:1994ph,Minkowski:2003jg,D:2003js,D:2004sj}.
In 1972, Bardeen, Gastmans and Lautrup \cite{bardeen} showed in
the frame of the SM and using the unitary gauge that the NCR is
infinite and therefore it is not a physical quantity. Later, in
the same year, S.Y. Lee \cite{sylee} working in the unitary gauge
considered the $\nu_l l'$ scattering and defined the NCR including
besides the usual terms, diagrams in which the photon is replaced
by a neutral gauge boson $Z$. In this way he obtained a finite,
although gauge dependent quantity \cite{lucio}. One of the
earliest analyses of the neutrino charge radius, in the context of
the general one-loop electromagnetic form factor of a fermion in
electroweak theories, was carried out in 1977 by Lee and Shrock
\cite{Lee:1977tib}. These authors working in the context of the SM
and using the linear R$_\xi$ gauge showed explicitly that the NCR
is not only infinite, but also gauge dependent. Lee and Shrock
showed in their paper how a full calculation, including not just
charge-radius terms, but also box diagrams (which could not be
considered to be corrections to the neutrino electromagnetic
vertex) combined together to yield a gauge-independent total
amplitude. Hence, in order to look for a definition of a physical
neutrino charge radius one has to consider other diagrams which
contribute to the total amplitude of the physical process $\nu_l
l' \rightarrow \nu_l l'$. The papers written by S.Y. Lee, and by
B.W. Lee and R.E. Shrock inspired many works in which finite and
gauge independent quantities, based on the NCR, were introduced by
considering the $\nu_l l'$ scattering
\cite{lucio,degrassi,Musolf:1990sa,B:2002pd,B:2002nw,B:2000hf,Nardi:2002ir,P:2002te,Hirsch:2002uv,Fujikawa:2003ww,P:2003rx,B:2004jr,P:2005cs}.
We want to end this paragraph pointying out the following.
Eventhough that it has been already shown that the neutrino charge
radius is an infinite and gauge dependent quantity in the frame of
the standard model (SM) when just the proper diagrams are taken
into account, it is possible to define a physical neutrino charge
radius by considering the $\nu_l l'$ scattering, which becomes a
finite and gauge independent quantity, independent of the lepton
$l'$ used to define it and also which only gets contribution from
the proper neutrino electromagnetic vertex
\cite{B:2000hf,B:2002pd,B:2002nw,P:2002te}. Discussions on the
experimental bounds on the NCR can be found, for example, in
Refs.\cite{Grifols:1986ed,Grifols:1989vi,Allen:1990xn,Mourao:1992ip,Salati:1993tf,Barranco:2007ea}

On the other hand, it is well known that any fermion may develop an
anapole moment $a$ \cite{zeldovich}. In particular, the neutrino,
even massless, may have an anapole moment $a_\nu$. Measurements of
the solar neutrino flux at Super-Kamiokande established that at
least one neutrino is not massless \cite{fukuda}. Besides, from
atmospheric and accelerator neutrino oscillations, we know that
there is a non-vanishing mass difference \cite{Michael:2006rx}. From
solar and reactor neutrino oscillations, we know that there is a
different non-vanishing mass difference \cite{:2008ee}. So, at least
two neutrinos are not massless. The neutrino anapole moment (NAM)
has been discussed in great detail in the literature
\cite{apenko,barroso,Rosado:2002rh,abak,gongora,czyz1,czyz2,rosado,CR:2002qx}.
Here, we only want to remark that in the frame of the standard model
\cite{weinberg} it is satisfied the relation $a_{\nu}^{SM} = \langle
r^2_{\nu} \rangle ^{SM}/6$ for a massless Dirac neutrino
\cite{Rosado:2002rh,rosado}. The relation between the charge form
factor and the anapole form factor for massless active (left-handed)
neutrinos is not model-dependent. In fact, it is a consequence of
having an effective vertex for the neutrino with the left-handed
chirality projector.

In this work, we study possible new physics effects on the NCR in
a model independent approach by using the effective Lagrangian
technique~\cite{BW,W}, which is an appropriate scheme to study
those processes that are suppressed or forbidden in the SM.
Motivated by the highly gauge dependent behavior of the charge and
the anapole form factors, $f_{Q}(q^2)$ and $f_{A}(q^2)$, within
the context of the SM, we will focus on those effects that could
be induced by a Yang--Mills sector possessing a richer gauge
structure than that of the dimension--four theory. To this end, we
will consider an effective electroweak Yang--Mills sector that
includes $SU_L(2)$--invariants of dimension higher than four. As
we will see below, there is only one dimension--six $SU_L(2)$
invariant that induces the $WW\gamma$ vertex and contributes at
the one--loop level to these form factors, $f_{Q}^{O_W}(q^2)$ and
$f_{A}^{O_W}(q^2)$, respectively. Hence, we can write
$f_{Q}^{eff}(q^2)=f_{Q}^{SM}(q^2)+f_{Q}^{O_W}(q^2)$ and
$f_{A}^{eff}(q^2)=f_{A}^{SM}(q^2)+f_{A}^{O_W}(q^2)$, where
$f_{Q}^{SM}(q^2)$ and $f_{A}^{SM}(q^2)$ are the standard model
form factors. We will show that, as a consequence of the $SU_L(2)$
symmetry, the dimension--six $WW\gamma$ vertex gives a
contribution which leads to manifest gauge independent expressions
for the $f_{Q}^{O_W}(q^2)$ and $f_{A}^{O_W}(q^2)$ form factors.
This result arises, in part, due to the fact that the
$SU_L(2)\times U_Y(1)$ invariants of dimension higher than four
are not affected by the gauge--fixing procedure used in the
dimension--four theory. As a consequence, in the context of
effective theories, fermionic form factors would be made of
vertices that are not affected by the gauge--fixing procedure,
which eventually would lead to gauge independent form
factors~\cite{NST}. Even more, we will show that it is possible to
express these form factors as finite functions of $q^2$ by
renormalizing them in the sense of effective field
theories~\cite{RMS}. These form factors get contribution at the
one loop level only from the proper neutrino electromagnetic
vertex. Besides, for a massless neutrino and as a consequence of
having an effective vertex for the neutrino with the left-handed
chirality projector the relation $f_{Q}^{O_W}(q^2)=q^2
f_{A}^{O_W}(q^2)$ is fulfilled. Hence, $f_{Q}^{eff}(q^2)=q^2
f_{A}^{eff}(q^2)$ and therefore we get the relation
$a_{\nu_l}^{eff} = \langle r^2_{\nu_l} \rangle^{eff} /6$, as in
the standard model.

This paper is organized as follows. In Sec.\ref{sm}, the structure
of the electromagnetic and anapole form factors of the neutrino in
the context of the SM is briefly discussed, mainly to fix our
notation. In Sec.\ref{cal}, the calculation of the electromagnetic
and anapole form factors of the neutrino in the context of an
effective Yang-Mills electroweak theory is presented. The gauge
independence of these form factors, as well as the possibility of
introducing a renormalization scheme beyond the Dyson's sense,
will be emphasized. Finally, in Sec.\ref{con} the conclusions are
presented.

\section{The electromagnetic structure of the $\bar{\nu}\nu \gamma$ vertex in the
SM}\label{sm}

In this section, we will present a schematic discussion on the main
features of the charge and anapole form factors of a massless Dirac
neutrino within the context of the SM \cite{weinberg} . We will take
advantage to introduce the notation and conventions that will be
used through the paper. For a massless left-handed neutrino the
matrix element of the electromagnetic current can be expressed in
terms of only one form factor $F(q^2)$ as
\begin{equation}
{\cal M}_\mu = i e F(q^2) \bar u_\nu (p') \gamma_\mu (1 - \gamma_5)
u_\nu (p).
\end{equation}
For a massless neutrino, this expression can easily be rewritten as
follows

\begin{equation}
{\cal M}_\mu = i e \bar u_\nu (p') \{ \gamma_\mu f_Q^{SM}(q^2) -
\gamma_\lambda \gamma_5 [{g^{\lambda}}_{\mu} q^2 - q^{\lambda}
q_{\mu}] f_A^{SM}(q^2) \} u_\nu (p)
\end{equation}

\noindent where\cite{marshak} $f_Q^{SM}(q^2) = F(q^2)$ and
$f_A^{SM}(q^2) = F(q^2)/q^2$ are the charge and anapole form factors
of the neutrino, respectively. $f_Q^{SM}(q^2)$ satisfies the
physical requirement:

\begin{equation}
f_Q^{SM}(0)=0.
\end{equation}
On the other hand, the NCR and the NAM are defined, respectively, by

\begin{equation}
\langle r^2_{\nu} \rangle = 6 \frac{\partial f_Q^{SM}(q^2)}{\partial
q^2} \mid_{q^2=0} \hspace{2mm} = 6 \frac{\partial F(q^2)}{\partial
q^2} \mid_{q^2=0},
\end{equation}

\noindent and

\begin{equation}
a_\nu = f_A^{SM}(0) = \frac{F(q^2)}{q^2} \mid_{q^2=0} \hspace{2mm} =
\frac{\partial F(q^2)}{\partial q^2} \mid_{q^2=0}.
\end{equation}

\noindent That is,

\begin{equation}
a_\nu^{SM} = \langle r^2_{\nu} \rangle^{SM} /6.
\end{equation}

\begin{figure}
\centering
\includegraphics[width=3.0in]{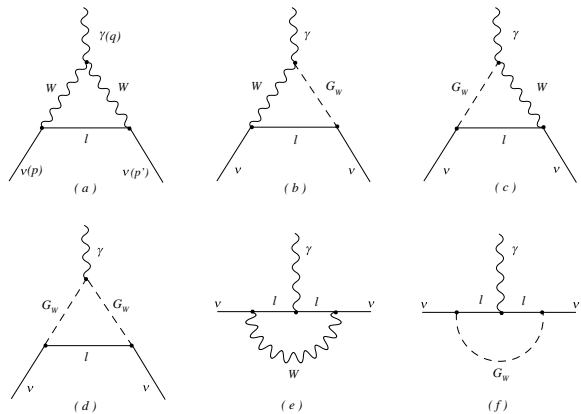}
\caption{\label{i} Proper diagrams contributing to the neutrino
charge radius in the standard model in the linear $R_\xi$ gauge.}
\end{figure}

\begin{figure}
\centering
\includegraphics[width=3.0in]{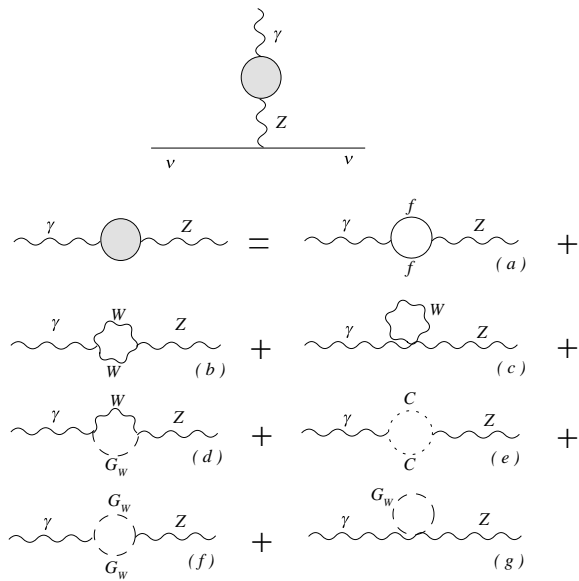}
\caption{\label{r} Reducible diagrams contributing to the neutrino
charge radius in the standard model in the linear $R_\xi$ gauge.}
\end{figure}

\section{Electromagnetic structure of a Dirac massless neutrino in an effective Yang-Mills theory}
\label{cal}

We now turn to calculating the one--loop contribution of an
anomalous (nonrenormalizable) $WW\gamma$ vertex to the charge and
anapole form factors of the SM Dirac neutrino. Next, we will
predict the corresponding charge radius by adopting an appropriate
renormalization scheme.

\subsection{The most general gauge structure of the $WW\gamma$
vertex} The gauge structure of the $WW\gamma$ vertex (and also of
$WWZ$) has been the subject of important attention in the
literature in diverse contexts. The one--loop radiative
corrections to the renormalizable vertex have been calculated in
the SM~\cite{WWgSM} and some of its extensions~\cite{WWgBSM}. Its
most general structure has been parametrized in a model
independent manner using the effective Lagrangian
approach~\cite{GG,HP,NT} and used in countless phenomenological
applications~\cite{DIVERSE}. One effective electroweak Lagrangian
can be constructed by adding to the dimension four Lagrangian all
the $SU_L(2)\times U_Y(1)$ invariants of dimension higher than
four, which may respect or no discrete transformations, such as
$P$, $T$, and $C$ or some combinations of them. The effective
Lagrangian can be written as
\begin{equation}
\label{el} {\cal L}_{eff}={\cal
L}_{SM}+\sum_{n=5}\sum_{i=1}^N\frac{\alpha_i}{\Lambda^{n-4}}{\cal
O}_i,
\end{equation}
where ${\cal L}_{SM}$ represents the SM Lagrangian and ${\cal
O}_i$ are $SU_(2)\times U_Y(1)$ invariants of dimension higher
than four. $\Lambda$ represents the new physics scale and the
$\alpha_i$ are unknown coefficients, which can be calculated once
the fundamental theory is known. The $SU_L(2)\times U_Y(1)$
structure of some of the ${\cal O}_i$ operators depend on the
mechanism responsible for the electroweak symmetry breaking. If
this breaking occurs through the Higgs mechanism, the
$SU_L(2)\times U_Y(1)$ symmetry is linearly realized through the
Higgs doublet. If this is not the case, the electroweak symmetry
is realized nonlinearly through the introduction of the matrix
field $U=exp(\sigma^a\phi^a/v)$ instead of the Higgs doublet.
Here, the $\phi^a(x)$ fields represent the pseudo Goldstone bosons
and $v$ is the Fermi scale. In this case, the effective Lagrangian
parametrizes new physics that are the responsible for the
electroweak symmetry breaking~\cite{NL}. In this paper, we will
focus on those type of effective interactions that are independent
of the mechanism responsible for the electroweak symmetry
breaking. As we will see below, this class of invariants induce
the nonrenormalizable structure of the $WW\gamma$ vertex, which is
dictated exclusively by the $SU_L(2)$ group.

From the known particles, the $W$ gauge boson, whose properties
would exhaustively be studied at the next generation of linear
colliders, is the one which possesses the richer collection of
electromagnetic form factors, as it is charged and has the highest
spin within the category of renormalizable theories. To our best
knowledge, Gaemers and Gounaris \cite{GG} derived initially 9 form
factors for the $WW\gamma$ vertex, but further on a careful analysis
carried out by Hagiwara--Peccei--Zeppenfeld--Hikasa \cite{HP} showed
that only 7 of these quantities are independent indeed. Subsequent
studies have confirmed these results \cite{BH}. These form factors
define the charge, the magnetic and electric dipole moments, the
magnetic and electric quadrupole moments, and the CP--even and
CP--odd anapole moments of this particle. The $WW\gamma$ vertex is
given by the following Lagrangian:
\begin{eqnarray}
\mathcal{L}_{WW\gamma}&=&ie\Big[(W^-_{\mu \nu}W^{+\mu}-W^+_{\mu
\nu}W^{-\mu})A^\nu +F_{\mu
\nu}W^{-\mu}W^{+\nu} \nonumber \\
&&+2\Delta \kappa_\gamma F_{\mu
\nu}W^{-\mu}W^{+\nu}+\tilde{\kappa}_\gamma \tilde{F}_{\mu
\nu}W^{-\mu} W^{+\nu}+\frac{\lambda_\gamma}{m^2_W}W^-_{\lambda
\mu}W^{+\mu}_{\ \ \nu}F^{\nu
\lambda}+\frac{\tilde{\lambda}_\gamma}{m^2_W}W^-_{\lambda
\mu}W^{+\mu}_{\ \ \nu}\tilde{F}^{\nu \lambda}\nonumber
\\
&&-\frac{i\tilde{a}^\gamma_W}{m^2_W}(W^+_{\lambda
\nu}W^{-\lambda}+W^-_{\nu \lambda}W^{+\lambda})\partial_\rho F^{\rho
\nu}-\frac{ia^\gamma_W}{m^2_W}(\tilde{W}^+_{\lambda
\nu}W^{-\lambda}+\tilde{W}^-_{\lambda \nu}W^{+\lambda})\partial_\rho
F^{\rho \nu} \Big],
\end{eqnarray}
where $F_{\mu \nu}=\partial_\mu A_\nu-\partial_\nu A_\mu$,
$\tilde{F}_{\mu \nu}=(1/2)\epsilon_{\mu \nu \alpha \beta}F^{\alpha
\beta}$, etc. While the CP--even $(\Delta \kappa_\gamma,
\lambda_\gamma)$ parameters define the magnetic dipole and
electric quadrupole moments, the CP--odd ones  $(\tilde{
\kappa}_\gamma, \tilde{\lambda}_\gamma)$ determine the electric
dipole and magnetic quadrupole moments of the $W$ gauge
boson~\cite{HP}. On the other hand, $a^\gamma_W$ and
$\tilde{a}^\gamma_W$ represent the CP--even and CP--odd anapole
moments of this particle. The only terms in the above equation
which are explicitly invariant under the electromagnetic $U_e(1)$
gauge group are those associated with the form factors
$\kappa_\gamma$ and $\tilde{\kappa}_\gamma$. However, the above
interactions arise from a more fundamental Lagrangian which is
invariant under the electroweak $SU_L(2)\times U_Y(1)$ gauge
group. In the linear realization of the group \cite{BW,W}, the
electroweak effective Lagrangian that induces
$\mathcal{L}_{WW\gamma}$ can be written in terms of $SU_L(2)\times
U_Y(1)$ invariants that comprise interactions of up to dimension
eight as follows:
\begin{eqnarray}
\mathcal{L}_{eff}&=&-\frac{1}{4}W^a_{\mu \nu}W^{a\mu
\nu}+\frac{\alpha_{WB}}{\Lambda^2}(\Phi^\dag W^{\mu
\nu}\Phi)B_{\mu
\nu}+\frac{\tilde{\alpha}_{WB}}{\Lambda^2}(\Phi^\dag W^{\mu
\nu}\Phi)\tilde{B}_{\mu \nu}\nonumber \\
&&+\frac{\alpha_W}{\Lambda^2}\frac{\epsilon_{abc}}{3!}W^a_{\lambda
\mu}W^{b\mu}_{\ \ \nu}W^{c\nu
\lambda}+\frac{\tilde{\alpha}_W}{\Lambda^2}\frac{\epsilon_{abc}}{3!}W^a_{\lambda
\mu}W^{b\mu}_{\ \ \nu}\tilde{W}^{c\nu \lambda}\nonumber \\
&&+\frac{\alpha_{WDB}}{\Lambda^4}\Big(i(\Phi^\dag W_{\nu \lambda}
D^\lambda \Phi)+H.c.\Big)\partial_\mu B^{\mu \nu}\nonumber \\
&&+\frac{\tilde{\alpha}_{WDB}}{\Lambda^4}\Big(i(\Phi^\dag
\tilde{W}_{\nu \lambda} D^\lambda \Phi)+H.c.\Big)\partial_\mu
B^{\mu \nu},
\end{eqnarray}
where the dimension--four Yang--Mills term has been included.
Here, $D^\lambda$ is the covariant derivative of the
$SU_L(2)\times U_Y(1)$ group and $W^a_{\mu \nu}$ ($W_{\mu
\nu}=\sigma^a W^a_{\mu \nu}/2$) and $B_{\mu \nu}$ are the
respective field tensors. In addition, $\Phi$ is the Higgs
doublet. The two dimension--eight operators were introduced in
order to generate the anapole moments of the $W$ boson. They are
made of the $SU_L(2)\times U_Y(1)$--invariant piece $(\Phi^\dag
W_{\nu \lambda} D^\lambda \Phi)$ \cite{LPTT} needed to generate
the $W$ current. These operators, which have not been considered
in the literature, can be eliminated using the equations of
motion, as must be. As for the dimension--six invariants, their
properties have widely been studied in the literature in diverse
contexts \cite{DIVERSE}.

\subsection{The anapole moment of the neutrino}

As already mentioned in Sec.\ref{sm}, a massless Dirac neutrino
possess only two electromagnetic form factors, which are however
not independent. These form factors are the charge, $f_Q(q^2)$,
and the anapole, $f_A(q^2)$, which satisfy the relation
$f_Q(q^2)=q^2f_A(q^2)$. The $WW\gamma$ vertex can contribute at
the one--loop level to these form factors through the
CP--conserving structures. However, in this work we will focus on
the contribution given by the CP--even $\lambda_\gamma$ component
of this vertex, which is governed exclusively by the $SU_L(2)$
gauge group. A comprehensive analysis for the off--shell vertex
$\bar{f}fV$, with $f$ an arbitrary fermion and $V=\gamma,Z$, will
be presented elsewhere~\cite{NST}. There are several good reasons
to consider the contribution given by the $SU_L(2)$--invariant
${\cal O}_W=(\epsilon_{abc}/3!)W^a_{\lambda \mu}W^{b\mu}_{\ \
\nu}W^{c\nu \lambda}$ to the neutrino form factors. In first place
is its pure Yang--Mills nature, which leads to the most general
structure for the gauge $WW\gamma$ vertex. This invariant
constitutes therefore a good theoretical instrument to studying
the gauge structure of electromagnetic form factors of elementary
particles. Another important reason to use this operator is, as we
will see below, that it leads to a NCR and NAM that are gauge
independent in manifest way. Also, due its nonrenormalizable
nature, ${\cal O}_W$ is necessarily generated at one--loop or
higher orders in any renormalizable theory, which means that it
maybe sensitive to new physics effects.

\begin{figure}
\centering
\includegraphics[width=3.0in]{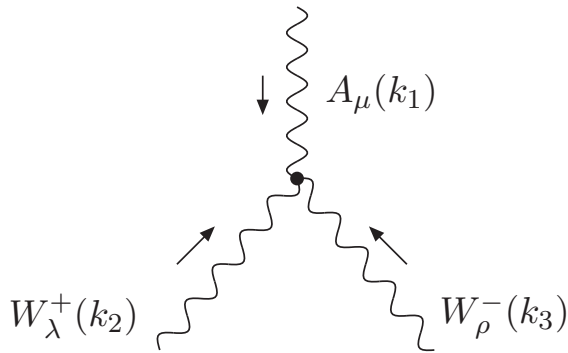}
\caption{\label{v} The trilinear $WW\gamma$ vertex.}
\end{figure}

Using the notation given in Fig.\ref{v}, the vertex function,
$(ie\lambda_\gamma/m^2_W)\Gamma_{\lambda \rho \mu}(k_1,k_2,k_3)$,
for the $\frac{ie\lambda_\gamma}{m^2_W}W^-_{\lambda \mu}W^{+\mu}_{\
\ \nu}F^{\nu \lambda}$ interaction can be written as:
\begin{equation}
\Gamma_{\lambda \rho \mu}(k_1,k_2,k_3)=(k_{3\alpha}g_{\beta
\rho}-k_{3\beta}g_{\alpha \rho})(k^\beta_2g_{\nu
\lambda}-k_{2\nu}\delta^\beta_\lambda)(k^\nu_1\delta^\alpha_\mu-k^\alpha_1\delta^\nu_\mu),
\end{equation}
which, as it is evident, satisfies the following simple Ward
identities:
\begin{eqnarray}
k^\mu_1\Gamma_{\lambda \rho \mu}(k_1,k_2,k_3)=0, \\
k^\lambda_2\Gamma_{\lambda \rho \mu}(k_1,k_2,k_3)=0, \\
k^\rho_3\Gamma_{\lambda \rho \mu}(k_1,k_2,k_3)=0.
\end{eqnarray}

We now turn to calculating the contribution of the above vertex to
the NAM. Since the ${\cal O}_W$ operator is not affected by the
gauge--fixing procedure of the dimension--four theory, there are no
contributions from ghost fields. Also, there are not contributions
from Goldstone bosons, as ${\cal O}_W$ does not depend on the
mechanism responsible for the electroweak symmetry breaking.
Accordingly, ${\cal O}_W$ only can contribute through the proper
diagram of Fig.\ref{i}$(a)$ and the self--energy diagrams given in
Figs.\ref{r}$(b)$ and \ref{r}$(c)$. The latter one being induced by
${\cal O}_W$ through the quartic $WWZ\gamma$ vertex. From all the
involved vertices, only the SM $WW\gamma$ one, which contributes
through the diagram in Fig.\ref{r}$(b)$, depends on the
gauge--fixing procedure. However, we have verified that due to the
above Ward identities, this diagram gives a vanishing contribution.
Also, we have encountered that there is no contribution from diagram
in Fig. \ref{r}$(c)$. So, the contribution to the NAM arises only
through the proper diagram given in Fig.\ref{i}$(a)$, whose vertices
are all independent of the gauge--fixing procedure. This means that
the only gauge dependence of the NAM could arise through the $W$
propagator, which in the $R_\xi$ gauge is given by
\begin{equation}
\Delta^{\mu \nu}=\frac{-i}{k^2-
m^2_W}\Big[g^{\mu\nu}-(\xi-1)\frac{k^\mu k^\mu }{k^2-\xi
m^2_W}\Big].
\end{equation}
However, it is clear that due to the Ward identities given above,
the longitudinal components of the $W$ propagators do not contribute
to the proper diagram given in Fig.\ref{i}$(a)$. As a consequence,
the amplitude for this diagram is manifestly gauge independent. This
amplitude is given by
\begin{equation}
{\cal M}^{O_W}_\mu
=\frac{ig^2\lambda_\gamma}{2m^2_W}\int\frac{d^Dk}{(2\pi)^D}\frac{P_R\gamma^\rho
\pFMSlash{k}\gamma^\lambda \Gamma_{\lambda \rho
\mu}}{[k^2-m^2_l][(k-p)^2-m^2_W][(k-p')^2-m^2_W]},
\end{equation}
where $m_l$ is the mass of the charge lepton to which is associated
the neutrino in consideration. Using the
Passarino--Veltman~\cite{PV} covariant decomposition, a direct
calculation leads to the following expressions for the charge and
anapole form factors:
\begin{eqnarray}
f_Q^{O_W}(q^2)&=&q^2f_A^{O_W}(q^2), \\
f_A^{O_W}(q^2)&=&\frac{\alpha \lambda_\gamma}{8\pi
m^2_W}\frac{1}{1-x_l}\Big[B_0(2)-x_lB_0(1)\Big],
\end{eqnarray}
where $B_0(1)=B_0(0,m^2_l,m^2_W)$ and $B_0(2)=B_0(q^2,m^2_W,m^2_W)$
are Passarino--Veltman two--point scalar functions. In addition, the
dimensionless variable $x_l=m^2_l/m^2_W$ was introduced. The above
expression was obtained after expressing a scalar $C_0$ three--point
function as a combination of $B_0$ functions~\cite{Stuart}. From the
above expressions, it is clear that the $f_Q^{O_W}(0)=0$ condition
is fulfilled. It is also evident that, although gauge independent,
the anapole form factor is divergent. In the next subsection, the
possibility of renormalizing this quantity within the context of
effective field theories will be explored.

\subsection{The neutrino charge radius}

The anapole form factor can be expressed in terms of elementary
functions as follows:
\begin{equation}
f_A^{O_W}(q^2)=\frac{\alpha \lambda_\gamma}{8\pi
m^2_W}\frac{1}{1-x_l}\Big[\Delta
(1-x_l)+2-x_l+(1-x_l)\log\Big(\frac{\mu^2}{m^2_W}\Big)-\frac{x^2_l}{1-x_l}\log(x_l)-g(x_q)\Big],
\end{equation}
where $\mu$ is the dimensional regularization scale and the
ultraviolet divergence is contained in:
\begin{equation}
\Delta =\frac{2}{4-D}-\gamma_E+\log(4\pi),
\end{equation}
with $\gamma_E$ the Euler's constant. In addition,
\begin{equation}
g(x_q)=\left\{\begin{array}{ll} 2\sqrt{\frac{4-x_q}{x_q}}\tan^{-1}\Big(\sqrt{\frac{x_q}{4-x_q}}\Big), & \textrm{if \ \ $x_q<4$ }\\
\sqrt{\frac{x_q-4}{x_q}}\Bigg[\log
\Bigg(\frac{1+\sqrt{\frac{x_q-4}{x_q}}}{1-\sqrt{\frac{x_q-4}{x_q}}}
\Bigg)-i\pi \Bigg] , & \textrm{if $x_q>4$}
\end{array} \right.
\end{equation}
where $x_q=q^2/m^2_W$. Following Refs.\cite{W,RMS}, the divergent
term in this amplitude can be absorbed by renormalizing the
coefficients of ${\cal L}_{eff}$ since it already contains all the
invariants allowed by the SM symmetry. The invariant needed to
absorb the divergence is
\begin{eqnarray}
\frac{\alpha_A}{\Lambda^2}(\bar{L}\gamma_\nu L)\partial_\mu B^{\mu
\nu},
\end{eqnarray}
where $L$ is the usual lepton doublet and $B^{\mu \nu}$ is the gauge
tensor associated with the $U_Y(1)$ group. In this way, the
divergence can be absorbed by the $\alpha_A$ parameter of the
effective Lagrangian. Here, we want to point out the following. The
divergency of Eq. (18) needs a counterterm in the effective
neutrino-photon vertex to be renormalized. The invariant of Eq. (21)
absorbs this divergence. However, after renormalization, the
contribution of the counterterm is not zero, but a finite value. As
a consequence, the final result is, in fact, the sum of two finite
contributions: one explicit depending on the effective $WW\gamma$
coupling and a second one coming from the counterterm proportional
to $\alpha_A$. The result quoted in this paper originates from the
first contribution. Using the $\overline{MS}$ renormalization scheme
with $\mu=\Lambda$, the renormalized anapole form factor can be
written as follows:
\begin{equation}
f_A^{O_W}(q^2)=\frac{\alpha \lambda_\gamma}{8\pi
m^2_W}\frac{f(x_q)}{1-x_l},
\end{equation}
where
\begin{equation}
f(x_q)=2-x_l+(1-x_l)\log\Big(\frac{\Lambda^2}{m^2_W}\Big)-\frac{x^2_l}{1-x_l}\log(x_l)-g(x_q).
\end{equation}

The anapole moment is the on--shell quantity $a_\nu=f_A(q^2=0)$. In
this limit, the loop function $f(q^2)$ takes the way
\begin{eqnarray}
f(0)&=&(1-x_l)\Big[2\log
\Big(\frac{\Lambda}{m_W}\Big)+1\Big]-\frac{x^2_l}{1-x_l}\log(x_l),\\
&\approx&2\log \Big(\frac{\Lambda}{m_W}\Big)+1,
\end{eqnarray}
which leads to the following expression for the additional
contribution in the frame of an effective theory to the neutrino
charge radius
\begin{eqnarray}
\langle r^2_{\nu} \rangle^{O_W}&=&\frac{3\alpha \lambda_\gamma}{4\pi
m^2_W}\Big[2\log\Big(\frac{\Lambda}{m_W}\Big)+1\Big],\\
&=&0.95\times 10^{-34}\ cm^2 \ \lambda_\gamma
\Big[2\log\Big(\frac{\Lambda}{m_W}\Big)+1\Big].
\end{eqnarray}

We now turn to discussing our results. In order to make predictions,
we need to assume some value for the $\lambda_\gamma$ parameter and
the new physics scale $\Lambda$. We will use the most recent
experimental limit on $\lambda_\gamma$ obtained by the D0
Collaboration~\cite{D02006}, namely, $-0.29<\lambda_\gamma <0.30$
for $\Lambda =2.0$ TeV. More recently, this Collaboration limited
the trilinear $WWZ$ vertex to $-0.17<\lambda_Z<0.21$ at the $95 \% $
C.L. and for the same value of $\Lambda$~\cite{D02007}. We will make
predictions assuming that $|\lambda_\gamma | \lsim 0.3$ and
$\Lambda=2.0$ TeV. Using these values, we obtain the following
additional contribution to the neutrino charge radius:
\begin{equation}
|\langle r^2_{\nu} \rangle^{O_W} | \lsim 2\times 10^{-34}\ cm^2.
\end{equation}

It is interesting to compare our result with the theoretical one
obtained, within the frame of the SM and by using the Pinch
Technique~\cite{PT}, in Ref.\cite{B:2000hf}. In this reference, it
was derived that $\langle r^2_{\nu_l} \rangle^{SM}=4.1, \, 2.4, \,
1.5 \times 10^{-33}\ cm^2$ for $l=e, \, \mu, \, \tau,$ respectively.
Hence we can conclude that our result is about one order of
magnitude lower than the results derived in Ref.\cite{B:2000hf}. On
the other hand, our result is of the same order of magnitude than
the new physics contribution derived in the context of the minimal
supersymmetric standard model~\cite{NCRSUSY}.

Finally, it is worthwhile to compare our result for the additional
contribution in the context of an effective theory to the neutrino
charge radius with some experimental bounds existing in the
literature. The upper bound $ \langle r^2_{\nu} \rangle^{exp}
<7\times 10^{-33}\ cm^2$ was derived from primordial nucleosynthesis
\cite{Grifols:1986ed}. In Ref.\cite{Grifols:1989vi}, the bound $
\langle r^2_{\nu} \rangle^{exp} <2\times 10^{-33}\ cm^2$ for the
charge radius of a right--handed neutrino was derived using
astrophysical information obtained from observational data on the SN
1987A. The bound $-2.74\times 10^{-32}<\langle r^2_{\nu}
\rangle^{exp} <4.88\times 10^{-33}\ cm^2$ was derived in
Ref.\cite{Allen:1990xn} from neutral--current neutrino reaction. In
Ref.\cite{Mourao:1992ip} the upper limit $ \langle r^2_{\nu}
\rangle<2\times 10^{-32}\ cm^2$ was derived from Kamiokande II and
Chlorine solar experiments. From the above considerations, we can
see that in general terms our result for $|\langle r^2_{\nu}
\rangle^{O_W} |$ is about one order of magnitude lower that the best
bounds derived so far.

\section{Conclusions}
\label{con}

Our aim in this paper has been to present the calculation of the
additional contribution, in the frame of an effective Yang-Mills
theory, to the charge and the anapole form factors,
$f_{Q}^{O_W}(q^2)$ and $f_{A}^{O_W}(q^2)$, for a massless Dirac
neutrino. These form factors get contribution at the one loop level
only from the proper neutrino electromagnetic vertex. We showed that
this vertex function is independent of the gauge--fixing parameter
and ultraviolet finite, as it can be renormalized within the
framework of the effective theory. We also showed that the relation
$f_{Q}^{O_W}(q^2)=q^2f_{A}^{O_W}(q^2)$ is fulfilled and hence we get
the relation $a_{\nu}^{O_W} = \langle r^2_{\nu} \rangle^{O_W} /6$.
Therefore, $a_{\nu}^{eff} = \langle r^2_{\nu} \rangle^{eff} /6$ as
in the standard model. This well-known relation between the charge
form factor and the anapole form factor for massless active
(left-handed) neutrinos does not depend on an specific model. This
relation is a consequence of having an effective vertex for the
neutrino with the left-handed chirality projector.

An interesting point, long discussed in the literature, is the gauge
independence and non-divergence of the neutrino charge radius. In
the SM, this issue is now clarified with a definition of this
quantity with all the required properties, which is furthermore
gauge independent and finite. In the model discussed by the present
paper, one should again treat these points with care, particularly
in the context of effective theories. The question of gauge
independence is solved favorably because only the transverse part of
the W-propagator appears. However, the divergency of Eq. (18) needs
a counterterm in the effective neutrino-photon vertex to be
renormalized. As it was stated in subsection III.C, the invariant of
Eq. (21) absorbs this divergence. However, after renormalization,
the contribution of the counterterm is not zero, but a finite value.
As a consequence, the final result is the sum of two finite
contributions: one explicit depending on the effective $WW\gamma$
coupling and a second one coming from the counterterm proportional
to $\alpha_A$. The result quoted in this work originates from the
first contribution.

Finally, we obtain $\langle r^2_{\nu} \rangle^{O_W} \approx
\frac{3}{4}(\alpha \lambda_{\gamma}/m_W^2)[2\log (
(\Lambda^2/M_W^2)+1]$. Using the recent D0 Collaboration
constraint on the $\lambda_\gamma$ parameter, we estimate the
value $|\langle r^2_{\nu} \rangle^{O_W} | \lsim 2\times 10^{-34}\
cm^2$, which is of the order of magnitude that may be expected in
theories beyond the SM, as supersymmetry, and  about one order of
magnitude lower than the current bounds.

\begin{center}
{\bf ACKNOWLEDGMENTS}
\end{center}

This work was supported in part by the {\it Consejo Nacional de
Ciencia y Tecnolog\'{\i}a} (CONACyT) and {\it Sistema Nacional de
Investigadores (SNI) de M\'exico}. A.R. would like to thank Prof.
R.E. Shrock for valuable comments.

%The authors would like to thank ... of this article for enlightening comments.

\end{document}